\documentclass[twocolumn]{aastex62}

\received{January 1, 2018}
\revised{January 7, 2018}
\accepted{\today}
\submitjournal{ApJ}

\shorttitle{Protostellar disk fragmentation}
\shortauthors{Sigalotti et al.}

\begin{document}

\title{From large-scale to protostellar disk fragmentation into close binary stars}

\correspondingauthor{Jaime Klapp}
\email{jaime.klapp@inin.gob.mx}

\author{Leonardo Di G. Sigalotti}
\author{Fidel Cruz}
\affil{{\'A}rea de F{\'{\i}}sica de Procesos Irreversibles, Departamento de
Ciencias B{\'a}sicas,\\
Universidad Aut{\'o}noma Metropolitana - Azcapotzalco (UAM-A),
Av. San Pablo 180, 02200 Ciudad de M{\'e}xico, Mexico}

\author{Ruslan Gabbasov}
\affil{Instituto de Ciencias B{\'a}sicas e Ingenier\'ias,
Universidad Aut{\'o}noma del Estado de Hidalgo (UAEH),\\
Ciudad Universitaria, Carretera Pachuca-Tulancingo km. 4.5 S/N, Colonia Carboneras,\\
Mineral de la Reforma, C.P. 42184, Hidalgo, Mexico}

\author{Jaime Klapp}
\affil{Instituto Nacional de Investigaciones Nucleares (ININ), Departamento 
de F{\'{\i}}sica\\ 
Carretera M{\'e}xico-Toluca km. 36.5, La Marquesa, 52750 Ocoyoacac, 
Estado de M{\'e}xico, M\'exico}
\affil{{\sc abacus}-Centro de Matem{\'a}tica Aplicada y C\'omputo de Alto Rendimiento,\\
Departamento de Matem{\'a}ticas, Centro de Investigaci{\'o}n y de Estudios Avanzados (Cinvestav-IPN),\\ 
Carretera M\'exico-Toluca km. 38.5, La Marquesa, 52740 Ocoyoacac,
Estado de M\'exico, M\'exico}

\author{Jos{\'e} Ram{\'{\i}}rez-Velasquez}
\affil{Instituto Venezolano de Investigaciones Cient{\'{\i}}ficas (IVIC), Centro de 
F{\'{\i}}sica,\\ 
Apartado Postal 20632, Caracas 1020A, Venezuela}
\affil{ESPOL Polytechnic University, Escuela Superior Polit\'ecnica del Litoral (ESPOL),\\
Physics Department, FCNM, Campus Gustavo Galindo km. 30.5, V{\'{\i}}a Perimetral,\\
P.O. Box 09-01-5863, Guayaquil, Ecuador}

\begin{abstract}

Recent observations of young stellar systems with the Atacama Large 
Millimeter/submillimeter Array (ALMA) and the Karl G. Jansky Very Large Array
(VLA) are helping to cement the idea that close companion stars form via 
fragmentation of a gravitationally unstable disk around a protostar early in 
the star formation process. As the disk grows in mass, it eventually becomes 
gravitationally unstable and fragments, forming one or more new protostars in 
orbit with the first at mean separations of 100 astronomical units (AU) or even
less. Here we report direct numerical calculations down to scales as small as 
$\sim 0.1$ AU, using a consistent Smoothed Particle Hydrodynamics (SPH) code,
that show the large-scale fragmentation of a cloud core into two protostars 
accompanied by small-scale fragmentation of their circumstellar disks. Our 
results demonstrate the two dominant mechanisms of star formation, where the 
disk forming around a protostar, which in turn results from the large-scale 
fragmentation of the cloud core, undergoes eccentric ($m=1$) fragmentation to 
produce a close binary. We generate two-dimensional emission maps and simulated 
ALMA 1.3 mm continuum images of the structure and fragmentation of the disks 
that can help explain the dynamical processes occurring within collapsing cloud
cores.

\end{abstract}


\section{Introduction} \label{sec:intro}

Stellar systems, consisting of two or more stars, are formed by two different 
mechanisms -- large-scale fragmentation of cloud cores of gas and dust during 
their early isothermal collapse phase \citep{Larson01,Fisher04}, leading to 
widely separated companion protostars [$\sim 1000$ astronomical units (AU) or 
larger], and small-scale fragmentation of their circumstellar disks early in 
the collapse due to gravitational instabilities \citep{Rodriguez05,Connelley08,
Pech10,Murillo13,Tobin13}, leading to bound systems separated by tens of AU. 
This scenario of star formation seems to be consistent with multiplicity surveys 
of the stellar population in nearby star-forming regions \citep{Connelley08,Kraus11,
Tobin16a}. Circumstellar disks are then expected to play an important role at 
closer separations, implying that many young stars with widely separated 
companions should also have a close companion \citep{Adams89,Andalib97}. New 
clues fitting this idea have been recently revealed by Karl G. Jansky Very Large
Array (VLA) and Atacama Large Millimeter/submillimeter Array (ALMA) observations 
of previously-unseen young protostars \citep{Tobin16a,Tobin16b}.

Numerical simulations of idealized protostellar disks with a central object have 
shown unstable behavior toward the development of long-wavelength, spiral shaped 
instabilities, with the eccentric mode \citep{Bonnell94,Woodward94}, when the 
disk-to-central-object mass ratio $M_{\rm d}/M_{\star}\geq 0.3$ \citep{Adams89}.
However, these models lack physical reality since protostellar disks form and
grow in mass within a collapsing environment as a result of the infall of
material from the outer cloud envelope \citep{Eisner12,Tobin12}. On the other
hand, if more protostars form within the same environment, the gravitational 
interaction among them may limit the disk size and affect the way angular
momentum is transferred. However, the disk fragmentation model has succeeded in
reproducing the constraints imposed by the observed statistical properties of
low-mass objects, such as low-mass binary systems, brown dwarfs, and even
planetary mass objects, which in turn have not yet been explained by means of
other formation mechanisms \citep{Stamatellos09}. Until now, no direct 
hydrodynamical collapse calculations have been reported in the literature that 
demonstrate the two dominant mechanisms for binary/multiple star formation mainly 
due to limited mass resolution. 

Here we present the results of high-resolution protostellar collapse simulations, 
using a consistent Smoothed Particle Hydrodynamics (SPH) code. The calculations
capture the structure and fragmentation of protostellar disks after the
large-scale fragmentation of a cloud core by spanning spatial
scales ranging from several thousands of AU to about 0.1 AU. We started the 
simulations from a gas sphere of uniform density and temperature representative 
of an idealized molecular cloud core, with parameters corresponding to the so-called 
``standard isothermal test case'' \citep{Burkert93}, coupled to a barotropic 
pressure-density relation to mimic the transition from the isothermal to the 
nonisothermal collapse phase \citep{Boss00}. We improve on mass resolution by 
allowing the calculations to work with unprecedented large numbers of particle 
neighbors, which translates into approximate second-order accuracy of the whole 
SPH discretization. Because of the lack of mass resolution, until now it was 
assumed that the collapse of the standard test case ends up with the formation 
of a wide binary with each core forming just a single object, since no 
fragmentation was seen in the simulations during the formation of the first 
protostars \citep{Kitsionas02,Arreaga07,Riaz14}. However, fragmentation of a
protostellar disk does not necessarily stop with the formation of a second 
protostar. Although our simulations do not show how the disks evolve much beyond
the initial disk fragmentation, present-day disk simulations predict that as fragmentation 
proceeds in high-mass accretion disks, new protostars can form at increasingly 
larger radii, with the process resulting in a chaotic multiple stellar system 
\citep{Krumholz09,Peters10,Bate18}. On the other hand, since gas accretion from
the outer envelope stops only when its angular momentum equals that of the
protostars, it is relatively easier for them to gain fresh gas through the
accretion disk, driving the system toward equal mass protostars \citep{Bate97}. 
The paper is organized as follows. In Section 2, we provide a brief description of 
the numerical methods along with the initial conditions employed for the collapse 
model calculations. The results of the simulations are described in Section 3 and 
the conclusions are given in Section 4.

\section{Numerical Methods} \label{sec:method}

\subsection{The SPH Scheme} \label{subsec:numerical}

The calculations of this paper were performed using a modified version of the SPH-based 
GADGET-2 code \citep{Springel05}, where the interpolation kernel was replaced by a 
Wendland C$^{4}$ function to allow support of large numbers of neighbors \citep{Wendland95,
Dehnen12}. Wendland functions have positive Fourier transforms and so they can work with
arbitrarily large numbers of neighbors without suffering from a pairing instability,
where particles come into close pairs and become less sensitive to small perturbations 
within the kernel support \citep{Dehnen12}. Moreover, Wendland kernels do not allow 
particle motion on a sub-resolution scale, maintaining a very regular particle distribution 
even in highly dynamical tests \citep{Rosswog15}. An improved scheme for the artificial 
viscosity relying on the method developed by \cite{Hu14} was also implemented, which uses 
a novel shock indicator based on the total time derivative of the velocity divergence with 
a limiter that applies the same weight to the velocity divergence and vorticity. This 
allows the artificial viscosity to distinguish true shocks from purely convergent flows and
discriminate between pre- and post-shocked regions. With this method viscous dissipation is 
effectively suppressed in subsonically convergent flows and in regions where the vorticity 
dominates over the velocity divergence, thus avoiding spurious angular momentum transport 
in the presence of vorticity. 

To ensure formal convergence and first-order (i.e., $C^{1}$) particle consistency of the SPH
equations, we adopt power-law dependences to set the smoothing length ($h$) and the number 
of neighbors ($n$) within the kernel support in terms of the total number of particles ($N$) 
\citep{Zhu15}. In particular, $C^{0}$ particle consistency (or first-order accuracy), i.e., 
satisfaction of the normalization condition of the kernel function in discrete form can only 
be achieved when $n$ is sufficiently large for which the finite SPH sum approximation 
approaches the continuous limit. This is consistent with the results of an error analysis of 
the SPH representation of the continuity and momentum equations, which show that particle 
consistency is lost due to zeroth-order errors that would persist when working with a fixed 
($=64$) number of neighbors even though $N\to\infty$ and $h\to 0$ \citep{Read10}. If $C^{0}$ 
particle consistency is achieved, then $C^{1}$ particle consistency is automatically ensured 
because of the symmetry of the kernel function. Out of the family of possible curves 
describing the dependence of $n$ on $N$, we choose the scaling relations $h\approx 7.68N^{-0.17}$ 
and $n\approx 7.61N^{0.503}$ to set the initial values of $h$ and $n$ in terms of $N$. 
These scalings were derived by requiring that $h/n\approx 2N^{-2/3}$, which accommodates
large numbers of neighbors for given $N$ while still keeping reasonably large values of $h$.
According to the parameterization $n\sim N^{1-3/\beta}$ given by \cite{Zhu15}, an exponent
of $\approx 0.503$ corresponds to $\beta\approx 6$, which is an intermediate value in the
interval $5\leq\beta\leq 7$. This interval is valid for low-order kernels as the Wendland 
C$^{4}$ function and $\beta\approx 6$ is appropriate for quasi-random particle distributions 
for which the particle approximation error goes as $n^{-1}$. As $N$ is increased, these 
relations comply with the asymptotic limits $N\to\infty$, $h\to 0$, and $n\to\infty$ with 
$n/N\to 0$ for full consistency of the SPH equations \citep{Rasio00}. The above scalings 
have important implications on the minimum resolvable mass $M_{\rm min}=nm$, where $m$ is 
the particle mass. Since $M_{\rm min}\sim h^{3}$ and $h\sim n^{-1/3}$, then 
$M_{\rm min}\sim n^{-1}$, implying improved mass resolution when the number of neighbors 
is increased. This feature makes a key difference with previous SPH collapse calculations 
since it allows resolution of small-scale structures in the flow through many orders of 
magnitude increase in density and pressure. Convergence and consistency testing on 
three-dimensional flow problems have demonstrated the second-order accuracy (i.e., the 
$C^{1}$-particle consistency) of the code \citep{Gabbasov17}.
\begin{figure*}
\plotone{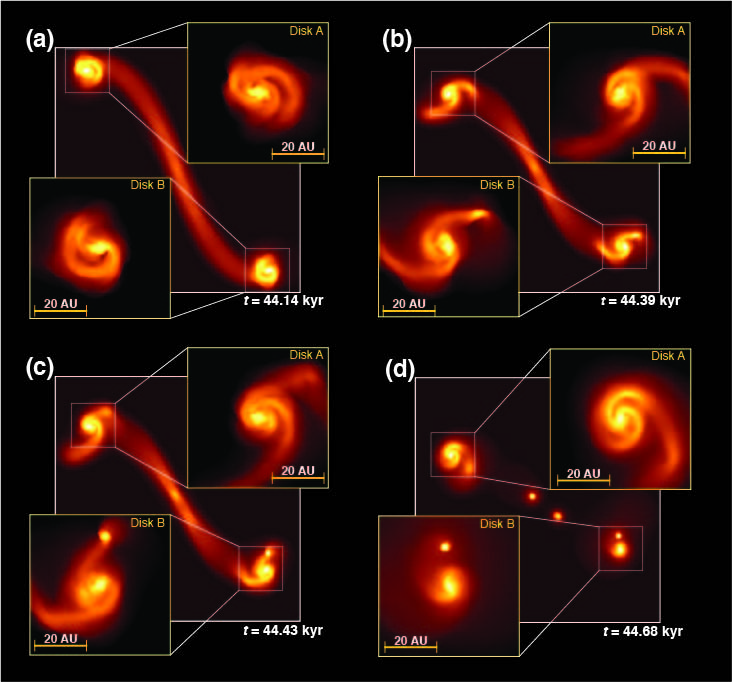}
\caption{Equatorial density evolution in a 266-AU region around the cloud center for 
model U2 after large-scale fragmentation into a binary protostar: (a) 44.14 kyr, (b) 
44.39 kyr, (c) 44.43 kyr, and (d) 44.68 kyr. The logarithm of the density $\rho$ is 
plotted. For (a), (b), (c), and (d) the maximum density is: $\log _{10}\rho _{\rm max}=-10.86$, 
$-10.63$, $-10.62$, and $-10.39$ ($\rho _{\rm max}$ in g cm$^{-3}$), respectively. The 
insets show a magnification of the binary cores and their whirling disks. The disks 
develop a two-armed spiral pattern associated with a linear growth of a gravitational 
instability. In (b) and (c), the gravitational instability in disk B enters a nonlinear 
growth phase and so one of its spiral arms is seen to fragment into a second protostar.
\label{fig:fig1}}
\end{figure*}
\subsection{Initial Conditions} \label{subsec:initial}

The initial conditions for protostellar collapse were chosen to be those of the standard 
isothermal test case \citep{Burkert93}, coupled to a barotropic pressure-density relation 
to mimic the transition from the isothermal to the nonisothermal collapse phase \citep{Boss00}.
The models were assumed to remain isothermal at an initial cloud temperature of 10 K up to a 
critical density $\rho _{\rm c}=5.0\times 10^{-12}$ g cm$^{-3}$, which separates the
isothermal from the nonisothermal collapse. Two highly resolved calculations were considered 
(models U1 and U2), which differed only in the total number of particles. Model U1 started 
with $N=2.4$ million particles and $n=12289$ neighbors, while model U2 had $N=4.8$ million 
particles and $n=17412$ neighbors. The particles were initially distributed in a glass 
configuration \citep{Couchman95} within a sphere of mass $M=1$ $M_{\odot}$, uniform density 
$\rho =3.82\times 10^{-18}$ g cm$^{-3}$, radius $R=0.016$ pc, and solid-body rotation 
$\omega =7.2\times 10^{-13}$ s$^{-1}$. The uniform-density background was perturbed azimuthally 
with a 10\%, $m=2$ variation. With this choice of the initial parameters, the ratios of the
thermal and rotational energies to the absolute value of the gravitational energy are, 
respectively, $\alpha\approx 0.26$ and $\beta\approx 0.16$. The models are close to virial 
equilibrium ($\alpha +\beta =1/2$), and apart from the assumption of uniform density, they 
could represent cloud cores that begin to infall after ambipolar diffusion has lessened the 
magnetic field support. The initial free-fall time is $t_{\rm ff}\approx 33.9$ kyr. Although 
the standard isothermal model is an idealization of a real cloud core, it provides a simple 
model from which to learn how nonaxisymmetric perturbations grow from a structureless medium.
In addition, when the transition from isothermal to adiabatic collapse is prolonged,
convergence is more difficult to attain and resolution requirements become significantly 
more demanding. These initial conditions are therefore suitable for testing the resolution 
performance of the code from several thousands of AU down to scales of 1 AU or less.

\section{Results}

Models U1 and U2 collapsed in a similar form. Figure 1 shows the density evolution of model
U2 at selected times after large-scale fragmentation into a binary protostar. A bar connecting 
two overdense fragments separated by $\approx 290$ AU is shown in Fig. 1a. Up to this
point the results of the simulations are very similar to those reported elsewhere in
the literature \citep{Kitsionas02,Arreaga07,Riaz14}, except that now the nascent binary
protostar is bridged by a centrally condensed, thicker bar. At this time, the binary cores 
are within the nonisothermal phase of contraction with maximum densities of $\approx 8.4$ 
orders of magnitude higher than the initial density. Spinning of the fragments causes the 
bar to fan out close to them and develop well-pronounced spiral arms (not shown in Fig. 1), 
which extend outward for about 350 AU. As the cores accrete low angular momentum mass from 
the bar and the outer spiral arms, their orbital separation decreases to around 191 AU and 
the bar dissipates (Figs. 1b-d). As a result of accretion, well-defined protostellar disks of 
size about 50 AU in diameter have formed around the primaries (disks A and B). After about 
250 years (Fig. 1b), the disks had increased in size and developed a strong two-arm spiral 
pattern. The larger insets in each frame show a magnification of the binary cores and their 
surrounding disks. Such spiral arms, other than providing the main source of angular momentum 
transfer, are a clear signature of self-gravitating disks. At this time, the gas in one of 
the spiral arms of disk B has become locally unstable and started to condense at a distance 
of 15 AU from the primary, forming a second protostar (Figs. 1b and c). By 44.68 kyr
(Fig. 1d), the bar has almost completely dissipated, leaving two linearly aligned condensations
of very low mass which move in opposite directions attracted by the larger and more massive
protostellar cores. At the time of Fig. 1d, the fragment formed from disk B has already 
completed a full revolution around the primary.
\begin{figure*}
\plotone{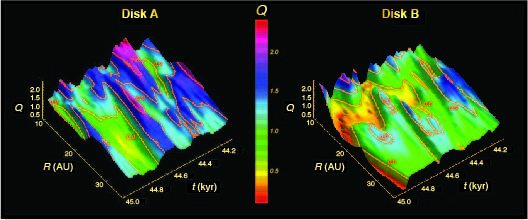}
\caption{Time variation of Toomre's $Q$ parameter along the radial structure of
disks A (left) and B (right). The color bar in the middle of both plots indicates the values 
of $Q$. Disk A shows values of $Q>0.6$ and hence it does not fragment. In contrast, $Q<0.6$ is 
achieved in one of the spiral arms of disk B as highlighted by the undulating yellow region
between 10 and 20 AU from the primary. This agrees with a recent revision of the Toomre's 
criterion which indicates that $Q<6$ in the spiral arms is a necessary condition for 
fragmentation of the disk \citep{Takahashi16}. As a result, the disk undergoes eccentric
($m=1$) fragmentation into a second protostar, forming a close binary.}
\end{figure*}
\subsection{The Toomre Parameter}

The standard reference for estimating the importance of gravitational instability
(GI) in young protostellar disks is the parameter $Q$, defined by \citep{Toomre64}
\begin{equation}
Q=\frac{c_{\rm s}\Omega _{\rm epi}}{\pi G\Sigma},
\end{equation}
where $c_{\rm s}$ is the sound speed, $\Omega _{\rm epi}$ is the epicyclic frequency
(which is equal to the angular velocity in a Keplerian disk), $G$ is the gravitational 
constant, and $\Sigma$ is the disk surface density. As $Q\to 1$, self-gravity in the 
disk becomes increasingly important and when $Q\approx 1$, the disk becomes unstable to
the growth of long-wavelength, spiral density waves. A recent revision of the Toomre's
criterion indicates that spiral arm formation occurs when $0.6<Q<1.0$, whereas
disk fragmentation due to nonlinear instability of the spiral arms sets in when
$Q<0.6$ \citep{Takahashi16}.

The behavior of self-gravitating disks is complex and depends on a range of disk
properties including size, surface density, temperature, and thermal physics. The
Toomre parameter at any position within the disk can be calculated using the
observationally more convenient expression given by \citep{Kratter16} 
\begin{equation}
Q=\frac{c_{\rm s}\Omega}{\pi G\Sigma}=f\frac{M_{\star}}{M_{d}}\frac{H}{R},
\end{equation}
where $M_{\star}$ is the mass of the central protostar, $M_{\rm d}$ is the disk mass,
$\Omega$ is the angular velocity, $H=c_{\rm s}/\Omega$ is the disk scale height, and
$f\approx 2$-4. For $Q>1$, the disk remains stable. As $Q\to 1$, self-gravity
dominates and the disk becomes marginally unstable to the growth of spiral density
waves. When $Q<1$, the disk may ultimately fragment due to nonlinear growth of the
GI. Figure 2 shows the time evolution of $Q$ along the radial structure of disks A
and B. The dynamics of both disks is dominated by the formation of spiral arms when
$0.6<Q<1.0$, whereas fragmentation of disk B occurs at $t\approx 44.39$ kyr when
$Q<0.6$ in one of the spiral arms (see Figs. 1b and c). This result is in very good
agreement with a recent revision of the condition for self-gravitational fragmentation
of protoplanetary disks \citep{Takahashi16}. Evidently, the condition $Q<0.6$ defines
the time and location within the disk where a GI sets in.

In order to evaluate the parameter $Q$ plotted in Fig. 2, we first calculate the
mass of the central core, $M_{\star}$, by summing up the mass of all particles lying 
within a sphere of radius 2 AU around the center of mass of the core. Similarly, the 
disk mass, $M_{\rm d}$, is calculated by summing up the mass of all particles contained 
within a cylindrical pillbox of radial extent 2 AU $\leq R\leq$ 50 AU and width 
$|z|\leq 20$ AU centered with the core. The upper frame in Fig. 3 shows the time 
evolution of $M_{\star}$ and $M_{\rm d}$ for disks A and B of model U2. Similar plots 
were obtained for the lower resolution case U1. For comparison, the solid red line 
depicts the mass of the gravitationally unstable region $M_{\rm GI}$ (i.e., the second 
fragment) in disk B as a function of time for which $Q<0.6$ according to Fig. 2. This 
mass grows steeply from $\sim 44.35$ kyr to $\sim 44.7$ kyr and then approximately 
linearly (at a much slower rate), reaching a mass of $\approx 0.02M_{\odot}$ by $44.98$ 
kyr. As the second protostar grows to a comparable mass to the primary (i.e., 
$\approx 0.02M_{\odot}$ against $\approx 0.034M_{\odot}$ for the primary), the disk
shrinks and eventually dissipates (Fig. 1d). During this stage, the secondary orbits
around the primary on a very short timescale compared to the stellar lifetime,
completing two orbits in less than 1 kyr by the time the simulations were terminated
(see Section 3.4 and Fig. 8 below for more details).
The bottom frame of Fig. 3 shows the time variation of the ratio $M_{\rm d}/M_{\star}$ 
for both disks. We note that disk B has $M_{\rm d}/M_{\star}>0.3$, as expected for 
the growth of the $m=1$ mode \citep{Adams89}. At $\approx 44.98$ kyr, 
$M_{\rm d}/M_{\star}\approx 0.7$ for disk B, whereas at this point disk A is not yet 
massive enough with values of $M_{\rm d}/M_{\star}$ that are barely above 0.3.

The scale height $H$ is calculated as a function of radius $R$ by assuming hydrostatic
equilibrium of the disk in the vertical $z$-direction. A rectangle of size $40\times 50$ 
AU$^{2}$ coinciding with the $\phi =0$ plane of the cylindrical pillbox is chosen on which
all particles within the box are projected. The rectangular domain
is first partitioned into 50 bins of radial width 1 AU and height 40 AU. The bins are
then subdivided into 40 square cells of area 1 AU$^{2}$, where cellwise values of the
density are calculated by averaging the contribution from all particles lying within
each cell. In this way, density profiles $\rho (z)$ are constructed in the $z$-direction
(i.e., along successive bins) for each radius and fitted to a Gaussian distribution of the
form \citep{Dong16}
\begin{equation}
	\rho (R,z)=\rho _{0}(R)\exp\left[\frac{-z^{2}}{2H^{2}(R)}\right],
\end{equation}
to determine $\rho _{0}(R)$ and $H(R)$, where $\rho _{0}(R)$ is the disk midplane
($z=0$) density. Part of this procedure is illustrated in Fig. 4, where the projected 
positions of particles on the $\phi =0$ plane of the pillbox are shown in the left panels 
for disks A and B. For clarity purposes only three bins bounded by the vertical red lines 
are depicted at selected radial positions. The panels on the right show the resulting 
Gaussian-like density distributions along the $z$-coordinate corresponding to the selected 
bins as indicated by the colored arrows. The maximum values of the distributions for each 
successive bin define the midplane density, $\rho _{0}(R)$, and the widths of the 
distributions define the scale height, $H(R)$. After formation of the protostellar disks, 
the values of $M_{\star}$, $M_{\rm d}$, $\rho _{0}(R)$, and $H(R)$ are determined after
each timestep and employed in Eq. (2) to calculate the parameter $Q$ displayed in Fig. 2 
as a function of disk radius and time.

\subsection{Two-dimensional Emission Maps}
\begin{figure}
\plotone{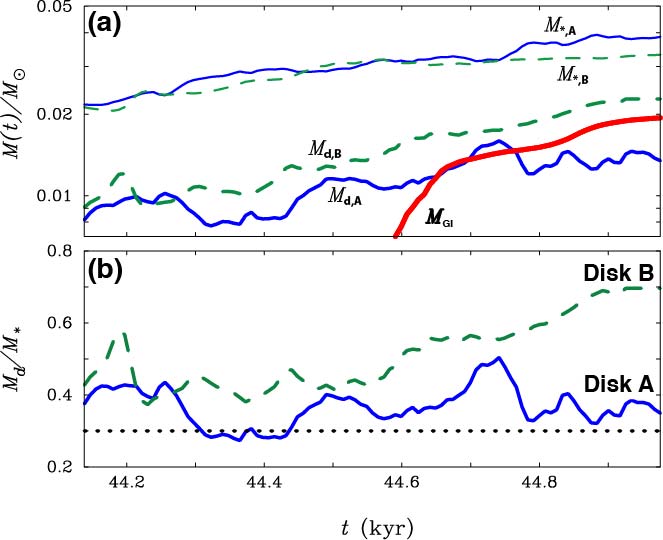}
\caption{(a) Time evolution of the masses of the primary protostars 
(thin solid blue and dashed green lines) and disks (thick solid blue and dashed green
lines) produced in model U2 for systems A and B. (b) Disk mass over central protostar
mass ratio, $M_{\rm d}/M_{\star}$, for disks A and B. The thick solid red line in (a) 
depicts the mass involved in the gravitationally unstable region within disk B as a 
function of time. The horizontal dotted line in (b) marks the analytically derived limit 
$M_{\rm d}/M_{\star}=0.3$ above which eccentric ($m=1$) disk fragmentation is expected 
to occur \citep{Adams89}. Disk B grows to $M_{\rm d}/M_{\star}\approx 0.7$, while disk A 
exhibits values of $M_{\rm d}/M_{\star}$ that are barely above the limit for eccentric 
fragmentation.}
\end{figure}
Two-dimensional (2D) emission maps of the protostellar disks of Fig. 1 can be generated
as follows. As a first step, we calculate the temperature for all particles by combining
the barotropic pressure-density relation with the ideal gas law to give
\begin{equation}
	T=T_{0}\left[1+\gamma\left(\frac{\rho}{\rho_{c}}\right)^{\gamma -1}\right],
\end{equation}
where $T_{0}=10$ K, $\rho_{c}=5.0\times 10^{-12}$ g cm$^{-3}$, and $\gamma=5/3$. At
temperatures below 100 K, a value of $\gamma=5/3$ is appropriate because the
rotational and vibrational degrees of freedom of molecular hydrogen are frozen out,
and so only translational degrees of freedom need be considered \citep{Boss00}. However,
precise knowledge of the dependence of temperature on density at the transition
from isothermal to nonisothermal collapse will require solving the radiative transfer
problem coupled to a fully self-consistent energy equation. This transition was
studied for the collapse of an initially centrally condensed core obeying a 
spherically-symmetric Gaussian density profile, using nonisothermal thermodynamics and 
solving the mean intensity equation in the Eddington approximation with detailed equations 
of state \citep{Boss00}. The collapse was seen to remain strictly isothermal
up to $\rho\sim 10^{-16}$ g cm$^{-3}$. At higher densities the collapse was near
isothermal, with the temperature rising very slowly from 10 K to $\sim 11.2$ K when
$\rho\sim 10^{-14}$ g cm$^{-3}$. At densities higher than this, the temperature was
seen to rise sharply with the density and the collapse became nonisothermal.
Therefore, a value of the critical density $\rho _{c}=1.0\times 10^{-14}$ g cm$^{-3}$
would be more representative of the near isothermal phase. However, a value of
$\rho _{c}=5.0\times 10^{-12}$ g cm$^{-3}$, while prolonging the isothermal phase of
collapse to higher densities, makes convergence to be more difficult to attain and
resolution requirements to be significantly more demanding. Larger and more massive
disks than those calculated here can therefore be obtained by anticipating the near 
isothermal phase using lower transition densities in barotropic collapse calculations. 
Figure 5 shows the projection on the disk midplane of the disk particle densities 
(normalized to the critical density $\rho _{c}=5\times 10^{-12}$ g cm$^{-3}$) and 
temperatures [calculated using Eq. (4)] as functions of radius for disks A and B at 
$\approx 44.39$ kyr. The prominent peaks in density and temperature at $R=0$ represent 
the central protostars, while those at $R\approx 18.2$ AU correspond to the secondary 
fragment produced in disk B. The upper portions of the density peaks are well above 
the line $\rho /\rho _{c}=1$, implying that the protostars are in a nonisothermal phase 
of collapse. The small peaks present at about 10 AU from the central protostar in Figs.
5a and c clearly indicates that by this time disk A is in the process of fragmenting 
into a new secondary as can be better seen from the emission maps of Fig. 6a.

The images displayed in Fig. 6 correspond to disk surface density and midplane 
temperature maps at selected times during the evolution of disks A and B. These images 
are generated from the numerical data by first drawing a cubic box of sides 56 AU that 
encloses the entire disk, where the geometrical center ($x=y=z=0$) and the $z=0$ plane 
of the box are made to coincide with the center-of-mass of the primary core and the 
aproximate equatorial plane of the disk, respectively. The box is divided into a number of
parallel $z=const.$ planes above and below the $z=0$ plane, with separations of 0.35 AU
between each other. Each successive plane is in turn partitioned into $150\times 150$ 
square pixels, with each pixel corresponding effectively to an area of 
$\approx 0.35\times 0.35$ AU$^{2}$. This results in a box composed of regular cubic cells 
of sides $\approx 0.35$ AU. A value of the density is assigned to each cell center
by simply averaging the density of all those particles within a cell. This procedure
permits drawing density profiles from cellwise densities along the $z$-direction over
the entire volume of the box, which are then fitted to a Gaussian distribution to
determine the disk midplane density, $\rho _{0}(x,y)$, and scale height, $H(x,y)$
(see Section 3.1). The disk surface density $\Sigma$ (in g cm$^{-2}$) is then calculated from 
the estimates of $H(x,y)$ using the expression $\Sigma =2\rho H$, where $\rho$ is the gas 
volume density \citep{Dong16}. The disk midplane temperature is finally calculated using 
Eq. (4) with $\rho$ replaced by $\rho _{0}(x,y)$ as determined from the maximum of the
Gaussian $z$-density profiles.
\begin{figure}
\plotone{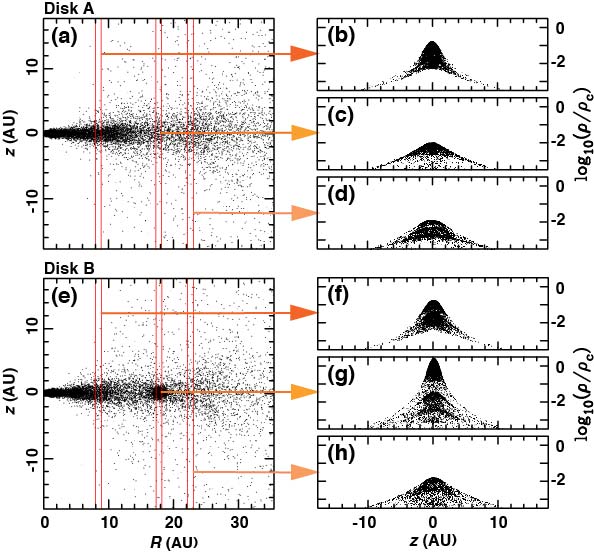}
\caption{(a) Projected particle positions on the $\phi =0$ plane of a cylindrical
pillbox of radius 50 AU and height 40 AU enclosing disk A. The vertical red lines binds three 
bins of radial width 1 AU at selected radii from the disk center. (b), (c), and (d) Gaussian-like 
density distributions corresponding to the bins in (a) as indicated by the colored arrows. 
(e) The same of (a) for disk B. (f), (g), and (h) The same of (b), (c), and (d) for disk B. 
The maximum and width of the distributions define the disk midplane density $\rho _{0}(R)$ and 
scale height $H(R)$ used in Eq. (3), respectively.}
\end{figure}
\begin{figure}
\plotone{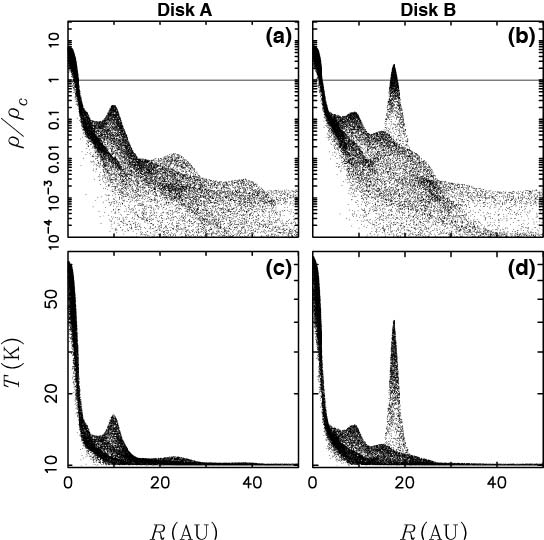}
\caption{Particle density distribution as a function of radius for (a) disk A
and (b) disk B. The horizontal solid line marks the transition from isothermal to
nonisothermal collapse. Also shown is the particle temperature distribution as a function of 
radius for (c) disk A and (d) disk B. The time shown in all plots is $t=44.39$ kyr. The density 
and temperature peaks centered at $R=0$ represent the primary protostars, while those centered 
at $R\approx 18.2$ AU correspond to the second fragment produced in disk B.}
\end{figure}
The radiative flux emerging from the disk surface is finally calculated using the
following relation \citep{Dong16}:
\begin{equation}
F(\Sigma _{1/2})=\frac{4\tau _{\rm P}\sigma T^{4}}{1+2\tau _{\rm P}+\frac{3}{2}
\tau _{\rm R}\tau _{\rm P}},
\end{equation}
where $\sigma$ is the Stefan-Boltzmann constant, $T$ is the disk midplane temperature,
and $\Sigma _{1/2}=\Sigma /2$ is the gas surface density from the midplane of the disk
to its surface. Equation (5) relates the emergent radiative flux from the disk surface
to the disk midplane temperature. Here $\tau _{\rm P}=\kappa _{\rm P}\Sigma _{1/2}$ and
$\tau _{\rm R}=\kappa _{\rm R}\Sigma _{1/2}$ are, respectively, the Planck and
Rosseland optical depths from the midplane to the disk surface, with $\kappa _{\rm P}$
and $\kappa _{\rm R}$ being the Planck and Rosseland mean opacities, respectively.
The Planck and Rosseland mean opacities are calculated from the opacity tables reported
by \cite{Semenov03}. We may see from Fig. 6 that disk A, which was less prone to 
fragmentation, shows clear signs of being forming a second protostar at a distance of
around 10 AU from the primary by $\approx 44.97$ kyr. This feature complies with
theoretical expectations of the role of disks in the formation of close stellar systems
\citep{Kratter16}.

\begin{figure*}
\plotone{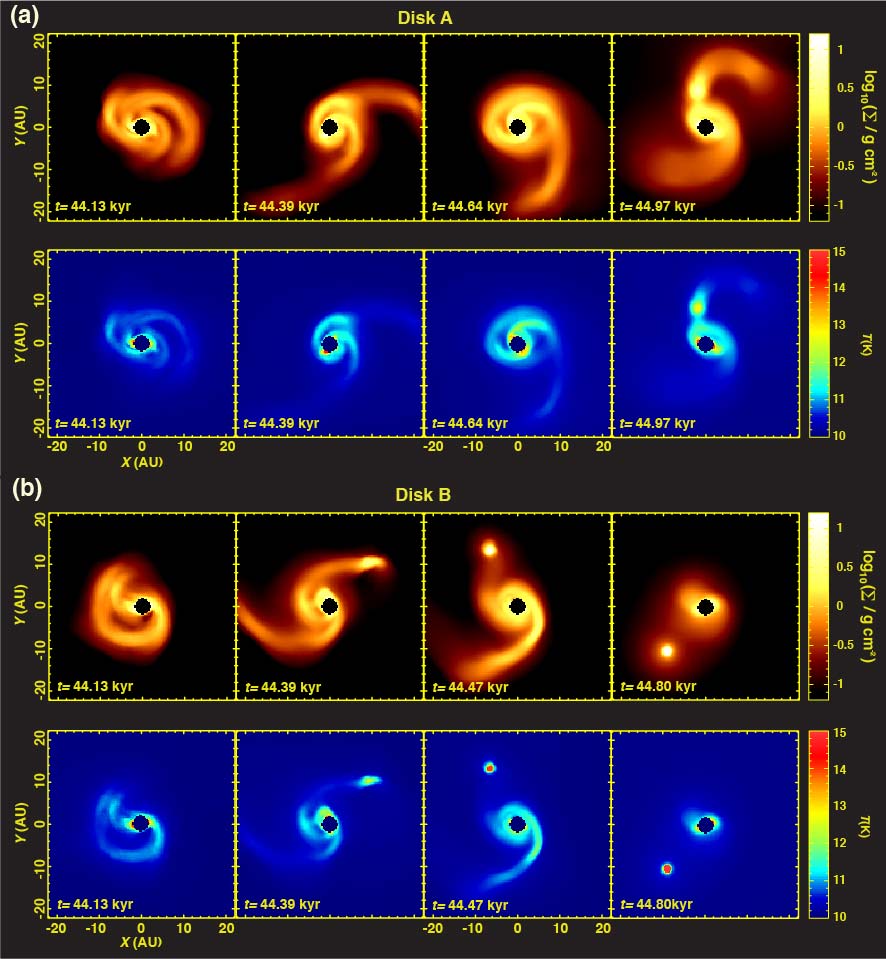}
\caption{(a) Gas surface density (top row) and midplane temperature (bottom row) 
of disk A viewed face-on at four different times. (b) The same of (a) for disk B. By 
$\approx 44.97$ kyr, when the calculation was terminated, disk A shows signs of fragmentation 
into a second protostar. The logarithm of the surface density $\Sigma$ (in g cm$^{-2}$) and 
the temperature $T$ (in K) are plotted. Each row of images uses the color bars on their right 
sides. In all images the central protostar is drawn with a black (surface density) and dark blue
(temperature) circle to increase the surface density and temperature contrasts.}
\end{figure*}

\subsection{ALMA Images}
\begin{figure*}
\plotone{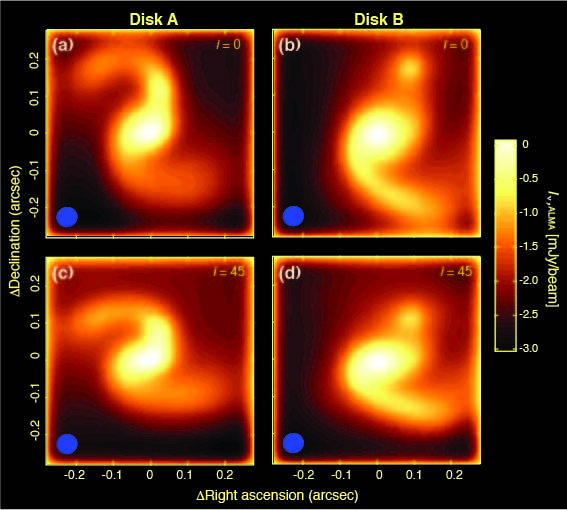}
\caption{ALMA 1.3 mm continuum images of (a) disk A at 44.97 kyr, (b) disk B at
44.47 kyr, both at a viewing angle of $0^{\circ}$, and (c) disk A at 44.97 kyr,
(d) disk B at 44.47 kyr, both at a viewing angle of $45^{\circ}$. The side
of the disks nearest the observer is oriented toward the bottom of each frame.
The source distance is assumed to be at 100 pc from the observer. The angular
resolution of each image is shown with the blue circle drawn in the lower left
corner, corresponding to a synthetic beam size of $0^{\prime\prime}.05$ (9 AU at 
this distance). The simulation images were produced with an integration time of 
1 hour. The color bar and numbers on the right indicate the flux intensity (in 
mJy beam$^{-1}$).}
\end{figure*}
Disks with $Q<0.6$ fragment very quickly on orbital timescales (i.e., within $\sim 10$
kyr). This process reduces the disk mass and stabilizes the system against further
activity \citep{Stamatellos11}. Therefore, the growth of the GI is a short-lived
phenomenon that is difficult to observe. In this sense it is useful to simulate ALMA
images to help interpret the observations. Figure 7 shows simulated ALMA images
of disks A and B at viewing angles of $0^{\circ}$ and $45^{\circ}$ and at two
different times during the evolution. These correspond to full resolution images obtained
by transforming the 2D emission maps of Fig. 6 with the aid of the {\tt simobserve} and
{\tt simanalyze} tools under Common Astronomy Software Application (CASA) for a wavelength
of 1.3 mm (230 GHz; ALMA band 6, configuration C43-8 consisting of a full array of 50
antennas of size 12 m). A reference distance of 100 pc from the source was assumed.
These images bear a remarkable resemblance to the ALMA images of the Class 0 triple
protostar L1448 IRS3 (consisting of a tight binary and a more widely separated tertiary)
as observed when the tertiary, which has the largest peak intensity in the system, is
removed \citep{Tobin16b}. However, compared to this protostellar system, the length
scale in our results for disk B is smaller by a factor of $\sim 4$.

\subsection{Orbital evolution of second protostar in disk B}

The orbital evolution of the second protostar around the primary in disk B
is shown in Fig. 8. The secondary emerges at a distance of $\approx 15$
AU from the primary protostar at about $44.38$ kyr, reaching a maximum separation
of $\approx 19$ AU by 44.43 kyr. During this stage the fragment accretes mass from the
disk at a high rate as shown by the steep increase of $M_{\rm GI}$, losing orbital
angular momentum during this stage and migrating to a minimum separation of
$\approx 10$ AU by $\approx 44.6$ kyr when it has already completed half of an
orbital period. Thereafter, as the disk dissipates the accretion rate declines and
the secondary moves away in its orbit until it reaches a maximum separation of
$\approx 18.2$ AU at $\approx 44.76$ kyr by the time it has completed a full revolution
about the primary. The secondary continues in its elliptic orbit and completes a
second period by $\approx 45$ kyr when the calculation is terminated. At this time,
the mass of the secondary has become comparable to the primary. During this second 
orbital period the minimum and maximum distances from the primary are $\approx 13.4$ and
$\approx 18.2$ AU, respectively. This gives an orbital period of $\approx 225$ years,
which corresponds to a much shorter timescale compared to the Class 0 lifetime
of $\sim 30$ kyr \citep{Andre00}. The eccentricity was found to decrease from
$\approx 0.36$ for the first period to $\approx 0.151$ for the second period,
suggesting that the secondary is approaching a more circular orbit as it comes close 
to the primary. However, a definite conclusion on the orbital stability of this 
double-star system (i.e., against mergers) cannot be reached because it will demand 
pursuing the calculation farther in time up to stellar densities.
\begin{figure}
\plotone{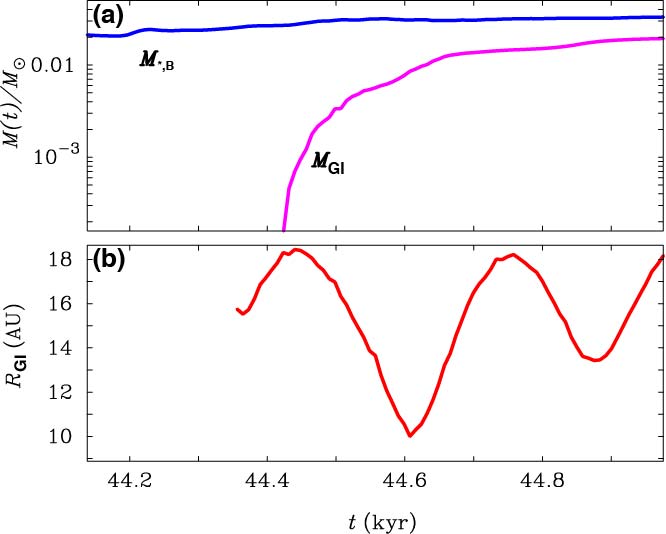}
\caption{(a) Time evolution of the masses of the primary $M_{\star}$ (solid
blue line) and secondary $M_{\rm GI}$ (solid magenta line) protostars in disk B. 
(b) Orbital evolution of the secondary core around the primary protostar. The curve 
displays the instantaneous position of the secondary after completion of two orbital 
periods.}
\end{figure}
\section{Discussion and Conclusions}

Fragmentation of the kind we have seen in our simulations should be a normal process in
the formation of stellar systems. However, at the stage of the evolution shown in Fig. 1d, 
slightly more than 10\% of the gas has been converted into protostars, and therefore the 
system is very much in its infancy. Beyond the epoch covered by our simulations, heating 
from the protostars may become strong enough to start dissociation of the H$_{2}$ available 
in the disk and so it will lose its ability to further fragment as most of the coolant is 
removed. As accretion from the cloud envelope continues, some protostars may become hot 
enough to initiate ionization of the atomic hydrogen, terminating the accretion. Other
than the host star irradiation, disks are also heated by two additional sources:
accretion energy and external radiation. However, for deeply embedded cores as may be
the case of Class 0 objects, the envelope may insulate the disk from external sources,
providing a somewhat lower temperature thermal bath \citep{Kratter16}.

Radiative hydrodynamics simulations of fragmenting disks were seen to reproduce the 
constraints set by the observed statistical properties of brown dwarfs and low-mass 
hydrogen-burning stars \citep{Stamatellos09}. More specifically, disk fragmentation can 
explain the shape of the mass distribution of low-mass stars, the lack of brown dwarfs as 
close companions to solar-mass stars, the presence of disks around brown dwarfs, the 
statistics of low-mass binary stars, and the formation of free-floating planetary objects. 
However, the main argument against the formation of brown dwarfs and low-mass stars by this 
mechanism is whether massive enough disks are actually realizable in nature. In view of the 
angular momentum content of their parental cores and the gas accretion from the cloud
envelope, the formation of such disks appears to be inevitable. On the other hand, massive 
($>0.2M_{\odot}$) and extended ($\sim 100$ AU) disks that do not fragment are
expected to dissipate on a viscous timescale (of order $\sim 1$ Myr) \citep{Stamatellos11}.
Simulations that ignore the effects of magnetic fields find that such disks form frequently 
in turbulent and/or rotating molecular cloud cores \citep{Attwood09}. Simulations of
protostellar disk formation including turbulence and magnetic fields also show that
protostellar disks usually do not extend over more than about 100 AU in size
\citep{Myers13,Joos13,Seifried15}. However, their rarity suggests that either they form and 
quickly fragment or their formation is suppressed by the effects of magnetic braking 
\citep{Mellon09}. Alternatively, if typical disk sizes are not larger than $\sim 100$ AU,
their non-detection could be a resolution issue. Using ideal magnetohydrodynamics (MHD),
\cite{Hennebelle08} finds that disk formation is suppressed if the magnetic field is
strong enough. On the other hand, if the magnetic field is misaligned, protostellar
disks may form but fragmentation is suppressed \citep{Hennebelle09,Commercon10}. However, 
resistive MHD calculations reveal that disk formation is possible because Ohmic 
dissipation dominates over other processes at relatively high densities 
($\sim 10^{12}$ cm$^{-3}$) \citep{Nakano02}. Further resistive MHD calculations by
\cite{Machida08} and \cite{Machida10} showed that disk formation and fragmentation after
the formation of the first core are possible, while \cite{Krasnopolsky10} found that
the formation of disks with sizes of $\sim 100$ AU will require resistivities of the
order of $10^{19}$ cm$^{2}$ s$^{-1}$.

In general, the masses and radii of protostellar disks formed at early stages of the
star formation process are much debated because of uncertainties in these
properties \citep{Stamatellos11}. The uncertainties arise because the disks around 
newly formed protostars (i.e., Class 0) are in almost all cases hidden within the 
infalling cloud envelope. Therefore, radiative transfer models are needed to distinguish 
the submillimeter and millimeter emission of the disk from the emission of the envelope. 
Although our simulations were not evolved much beyond the initial disk fragmentation,
comparison with observations indicates that many detected disks around Class I
and Class II objects may be remnants of Class 0 disks that have fragmented at a
very early stage. Disks around young protostars ($\sim 40$ kyr) with masses and
sizes consistent with the masses (0.01--0.03$M_{\odot}$) and sizes (20--50 AU)
predicted by our simulations have been observed \citep{Rodriguez05,Eisner12}.

Our results demonstrate that protostellar disks can experience gravitational 
instability and fragmentation at very young ages, leading to the formation of 
hierarchical multiple systems as was recently detected by ALMA observations 
\citep{Tobin16b}. The advent of mathematically consistent resistive MHD, 
radiative hydrodynamics models, together with the discovery of larger samples of Class 
0 objects with unstable disks using facilities such as ALMA, will help to assess the 
contribution of disk fragmentation to the production of close binary/multiple stars.

\acknowledgments

We acknowledge support from the {\sc abacus}-Centro de Matem\'atica Aplicada y 
C\'omputo de Alto Rendimiento of Cinvestav-IPN under CONACYT grant EDOMEX-2011-C01-165873.
The calculations of this paper were performed using the {\sc abacus} supercomputer
facilities.

\end{document}